\documentstyle[12pt]{article}
\begin{document}

\title{JAMMING TRANSITION WITH FLUCTUATIONS OF CHARACTERISTIC
ACCELERATION/BRAKING TIME WITHIN LORENTZ MODEL}

\author{Alexei Khomenko, Dmitrii Kharchenko and Olga Yushchenko\\
Department of Physical Electronics, Sumy State University,\\
2, Rimskii-Korsakov Str., 40007 Sumy, Ukraine,\\
{\it khom@phe.sumdu.edu.ua, kpe@sumdu.edu.ua}}
\maketitle
\begin{abstract}
Jamming transition in traffic flow
(between free and jammed traffic) for homogeneous car
following model has been investigated taking into account
 fluctuations of characteristic acceleration/braking time. These
fluctuations are defined by Ornstein-Uhlenbeck process. The
behaviour of the most probable deviation of headway from its
optimal value has been studied and phase diagram of the
system has been calculated for supercritical and subcritical
regimes of jam formation. It has been found that for the first
regime the fluctuations of characteristic acceleration/braking time
result in coexistence of free moving and jammed traffic,
that is typical for the first-order phase transition,
and in appearance of two steady states for the second mode.
These states correspond to non-zero values of headway
deviation at which the formation of jam and congested traffic
are possible. Using phase-plain portraits method the kinetics of the
system transitions has been analyzed for different domains
of the phase diagram for both regimes.
\end{abstract}

\section{Introduction}

Nowadays traffic problems attract considerable attention.
For a study of the traffic jams formation problem thermodynamic
\cite{1}, stochastic \cite{2}, and some hydrodynamic and
kinetic theories \cite{3,4} are used. These theories are based
on car-following model \cite{3,4}, Maxwell model \cite{5},
and cellular automaton model \cite{3,6}. Within the
framework of thermodynamic approach jamming transition is
represented as nonequilibrium first-order phase transition.
The mentioned method describes the deterministic picture
of traffic flow. Stochastic theory which is based on the master
equation allows to find a stationary density of spatially-temporal
distribution of traffic jams. However, the fluctuations
influence of the parameters  characterizing the system is
not finally studied. It is known that fluctuations do not only
play a role of the trigger for the phase transition but also lead to
essentially new system behaviour, and are a major cause for
system's self-organization \cite{7,8}. Therefore, at describing
the jamming transition the estimation of fluctuations influence
of the dynamical car characteristics is rational. This is
achieved within the framework of synergetic approach by the most
natural way.

One of the simplest schemes that describe the self-organization
process is the Lorentz system \cite{8,9}. At first time it was
offered for the atmospheric phenomena description and then
used for physics, chemistry, biology, sociology problems and
so on. Recently synergetic scheme was proposed to describe
the jamming transition in traffic flow \cite{10}. In our work
offered in Ref. \cite{10} the synergetic approach will be developed
taking into account internal fluctuations of characteristic
acceleration/braking time. This parameter is a car characteristic,
it indicates a time necessary for car to reach the characteristic
velocity and plays a key role at the traffic jam formation.
We will show that internal fluctuations, which have additive
noise meaning, lead to the complication of traffic flow behaviour. In this
work the stationary regime of such system will be considered and influence
of the characteristic acceleration/braking time  in the most
probable headway deviation from its optimal value will be studied.
As it will be shown this headway deviation characterizes a phase
transition into the state that has traffic jam meaning.

\section{Basic equations}

On the basis of the car following model for one-line highway it is
possible to show that dissipative dynamics of the homogeneous traffic flow
can be represented within the framework of the Lorentz scheme \cite{10}. To
describe the jamming transition we  use the synergetic concept of
phase transition, which is realized as a result of mutually coordinated
behaviour of three freedom degrees: an order parameter, a conjugate field,
and a control parameter. In traffic flow the roles of these quantities are
played by the absolute value of headway deviation 
between vehicular from a safety distance $h$
\begin{equation}
\eta \equiv |{\rm \Delta} x - h|;\label{1}
\end{equation}
\noindent
by the deviation of velocity of the  $\eta$ variation from optimal value
$h/t_0 - V$
\begin{equation}
v \equiv \Delta \dot {x} - h / t_{0} + V,\label{2}
\end{equation}
\noindent
where $t_0$ is the nominal time lag, $V$ is the actual value of the
velocity, and by acceleration/braking time $\tau$, respectively.
Here $x$ is the vehicle coordinate.

Let us consider the simplest homogeneous system with goal
of definition of time dependencies  $\eta(t)$, $v(t)$,
$\tau(t)$. To achieve this  we will use a phenomenological
approach. In the equations of the motion it is assumed
 that in the autonomous mode the evolution of the
quantities $\eta $, $v$ and $\tau$ has dissipative character
and their relaxation to equilibrium values is described
by the Debye equation. Besides a Le Chatelier principle
has a great importance: since the decrease of
acceleration/baking time  $\tau$ assists to the formation of
stable traffic flow, the headway deviation  $\eta $ and its velocity
deviation $v$ should vary so that to prevent the growth of
$\tau$, and as a consequence to impede a jam formation.
Also the essentially important role is played by positive
feedback of values $\eta $ and $\tau $ on $v$. Namely the
availability of this feedback is a reason for
self-organization that leads to traffic jam formation.

The Lorentz system takes into account the mentioned above circumstances by
the simplest way. Taking under consideration fluctuational addition
it is determined as follows
\begin{eqnarray}
\dot\eta  &=& -\frac{\eta}{t_{\eta}} + v,\label{3}\\
\dot v &=& - \frac{v}{t_v} + g_v \tau \eta ,\label{4}\\
\dot\tau & =& \frac{\tau_0 - \tau}{t_{\tau}} - g_{\tau} \eta
v + \lambda(t).\label{5}
\end{eqnarray}

\noindent Here the dot means the time differentiation; $t_{\eta}$,
$t_{v}$ and $t_{\tau}$ are the appropriate relaxation times;
$g_{v}$ and $g_{\tau}$ are the positive constants.

System (\ref{3})--(\ref{5}) represents a base of the self-con\-sis\-tent
description for the car-following model.
Here the first terms on the right-hand side of the equations
describe the relaxation of each quantity to an equilibrium
value. In equation (\ref{3}) the last term is a usual addition. It is
easy to see from this equation that in the stationary state headway
deviation is proportional to the velocity deviation.

The second term on the right-hand side of Eq. (\ref{4}) describes the positive
feedback of headway deviation $\eta $ and acceleration/braking time $\tau $
 on velocity deviation $v$. This feedback leads to the increase of $v$
and is the reason for traffic jam formation.

Equation (\ref{5})  differs from (\ref{3}), (\ref{4}) because
relaxation of $\tau $ occurs not to zero but to finite value  $\tau_0$
which represents the time necessary for automobile to reach a
characteristic velocity (the car property). Minus sign before
the last term on the right-hand
side of  equation (\ref{5}) may be considered as a demonstration
of the Le Chatelier principle.

Quantity  $\lambda(t)$ represents an influence of fluctuations
of the characteristic acceleration/braking time and is defined
as Ornstein-Uhlenbeck process:
\begin{eqnarray}
\left\langle{\lambda(t)}\right\rangle &=& 0,\nonumber\\
\left\langle{\lambda(t)\lambda(t^{\prime})}
\right\rangle& \equiv& C(t,t^{\prime}) =
\frac{\delta^2}{\tau_{\lambda}} e^{-\frac{\left|t-t^{\prime}
\right|}{\tau_{\lambda}}},\label{6}
\end{eqnarray}
\noindent
where $\delta^{2}$ is the noise intensity, $\tau_{\lambda}$ is the
correlation time of the process  $\lambda(t)$.

Within the framework of the mentioned parametrization the formation of
traffic jams is represented as a result of spontaneous headway and
velocity deviations if the characteristic acceleration/braking time exceeds
a critical value. It is reflected by the appearance of the minimum of the
effective potential which corresponds to the stationary value of the
headway deviation $\eta_{0}$ \cite{10}. Therefore  we will
be interested in $\eta $ evolution further.

In the general case system (\ref{3})--(\ref{5}) have no analytical
solution therefore we should use the following approximation:
\begin{equation}
t_{\eta} \gg t_{\tau},\quad
t_{\eta} \approx t_v.\label{7}
\end{equation}
\noindent
This condition implies that in the course of evolution the
acceleration/braking time $\tau$ is coordinated by variation
of the headway and velocity deviations. Owing to this
condition in equation (\ref{5}) a small parameter can be
eliminated that allows us to assume
$t_{\tau}\dot\tau\approx0$. As a result we derive expression for
the control parameter in the following form
\begin{equation}
\tau=\tau_0-g_{\tau}t_{\tau}\eta v + t_{\tau}\lambda(t).\label{8}
\end{equation}
\noindent
To form simpler system let us reduce initial one to the one-parameter model.
For that it is necessary to express $v$ and $\tau$ via $\eta$. Equation for
$\dot v$ is determined by differentiating with respect to
time equation for the velocity deviation $v$
obtained from Eq.(\ref{3}). Substituting expressions for $v, \dot v$,
and equation (\ref{8}) into (\ref{4}), and
introducing measure scales $t_{\eta}$,
$\eta_m=(g_v g_{\tau}t_{\tau}t_{\eta})^{-1/2}$, $v_m=t_{\eta}^{-3/2}
(g_v g_{\tau}t_{\tau})^{-1/2}$, $\tau_c=(g_v t_{\eta}^2)^{-1}$, and
$g_v t_{\tau}t_{\eta}^2$ for time, headway
deviation  $\eta$, velocity deviation $v$, acceleration/braking time
$\tau$, and for the noise of the characteristic acceleration/braking
time $\tau_0$, respectively, we get:
\begin{equation}
\ddot\eta+\dot\eta(1+\sigma+\eta^2)=\eta(\varepsilon-\sigma)-\eta^3
+\eta \lambda(t).\label{9}
\end{equation}
\noindent
Here the denotations are introduced:
\begin{equation}
\sigma\equiv\frac{t_{\eta}}{t_v},\quad
\varepsilon\equiv\frac{\tau_0}{\tau_c}.\label{10}
\end{equation}

Obtained expression allows us to write the evolution equation
in the canonical form of the motion equation for the nonlinear
stochastic oscillator of the van der Pole generator type
\begin{equation}
\ddot\eta+\gamma(\eta)\dot\eta= f(\eta)+g(\eta)\lambda(t),\label{11}
\end{equation}
\noindent
where
\begin{eqnarray}
\gamma(\eta)&=&1+\sigma+\eta^2,\nonumber\\
f(\eta)&=&\eta(\varepsilon-\sigma)-\eta^3,\label{12}\\
g(\eta)&=&\eta.\nonumber
\end{eqnarray}
\nonumber
Note that equation (\ref{11}) takes into account reactive
behaviour of the system. As is known the task of the statistical
physics is solved using distribution function $P(\dot\eta,\eta,t)$
that represents probability density of the
availability of the corresponding values of headway deviation $\eta$
and its rate of change $\dot\eta$ at a given instant of time  $t$.
Since jam in the car flow is defined by headway  $\eta$ and
time $t$ we have to consider projection of the distribution
function in the half-space ($\eta, t$). For that the
kinetic equation for $P(\dot\eta,\eta,t)$ is reduced to the Fokker-Planck
equation with respect to $P(\eta,t)$ function.

\section{Langevin and Fokker-Planck equations}

For derivation of the evolution equation for the probability
density $P(\dot\eta,\eta,t)$ we use continuity equation
for function $\rho(\dot\eta,\eta,t)$ which is
connected with $P$ by following equality $P(\dot\eta,\eta,t)=
\langle\rho(\dot\eta,\eta,t)\rangle_{\lambda}$.
Here $\langle \rangle_{\lambda}$ stands for averaging over noise $\lambda$.
Continuity equation is constructed by standard manner:
\begin{equation}
\left(\frac{\partial}{\partial t}+ \hat{L}(\dot\eta,\eta)\right)
\rho(\dot\eta,\eta,t) = -g(\eta)\lambda(t)
\frac{\partial}{\partial \dot\eta} \rho(\dot\eta,\eta,t),\label{13}
\end{equation}
\noindent
where the operator 
\begin{equation}
\hat{L}(\dot\eta,\eta)=-\gamma(\eta)\frac{\partial}{\partial\dot\eta}
\dot\eta + \dot\eta \frac{\partial}{\partial\eta}+ f(\eta)
\frac{\partial}{\partial\dot\eta}\label{14}
\end{equation}
\noindent
is introduced. Averaging equation (\ref{13}) 
and using decomposition technique
in cumulants \cite{11,12} we obtain kinetic equation for $P$:
\begin{equation}
\left\{ \frac{\partial}{\partial t}+\hat{L}(\dot\eta,\eta) \right\}
P(\dot\eta,\eta,t)=\hat{\Lambda}(\dot\eta,\eta,t)P(\dot\eta,\eta,t).\label{15}
\end{equation}
\noindent
Here
\begin{equation}
\hat{\Lambda}(\dot\eta,\eta,t) = g(\eta)\frac{\partial}{\partial\dot\eta}
\sum\limits_{k=0}^{\infty} {C^{(k)} \hat{L}^{(k)}(\dot\eta,\eta)},\label{16}
\end{equation}
\noindent
and moments of the correlation function are defined as follows:
\begin{equation}
C^{(k)}(t)=\frac{1}{k!} \int\limits_0^{\infty}\tau_{\lambda}^k
C(t,t-\tau_{\lambda})d\tau_{\lambda}.\label{17}
\end{equation}
\noindent
Operators $\hat{L}^{(k)}$ are defined by recurrent formula
\begin{equation}
\hat{L}^{(k)}= {\left[\hat{L}^{(k-1},\hat{L}\right]},\quad
\hat{L}^{(0)}=g(\eta)\frac{\partial}{\partial\dot\eta}.\label{18}
\end{equation}
\noindent
Here $[\hat{A},\hat{B}]=\hat{A}\hat{B}-\hat{B}\hat{A}$
is the denotation for the commutator .

To pass into the half-space ($\eta, t$) the distribution function
moments are used
\begin{equation}
P_n(\eta,t)=\int\limits_0^{\infty} (\dot\eta)^n P(\dot\eta,\eta,t)d\dot\eta.
\label{19}
\end{equation}
\noindent
Namely these moments give us an opportunity to find an effective Fokker-Planck
equation. Taking into account components of integer orders, instead of
(\ref{15}) according to \cite{13}, we get
\begin{equation}
\frac{\partial}{\partial t}P(\eta,t)=-\frac{\partial}{\partial\eta}
D^{(1)}(\eta) P(\eta,t)+ \frac{\partial^2}{\partial\eta^2}
D^{(2)}(\eta)P(\eta,t),\label{20}
\end{equation}
\noindent
where
\begin{eqnarray}
D^{(1)}(\eta)&=&\frac{f(\eta)+C^{(1)}\partial g^2(\eta)/\partial\eta}
{\gamma(\eta)}+\frac{C^{(0)}g^{2}(\eta)\nabla(\gamma^{-1}(\eta))}
{\gamma(\eta)},\label{21}\\
D^{(2)}(\eta)&=&\frac{g^2(\eta)}{\gamma^2(\eta)}\label{22}
\end{eqnarray}
\noindent
are the drift and diffusion coefficients, respectively.

Equation (\ref{20}) corresponds to the Langevin equation governing the
$\eta$ evolution
\begin{equation}
\dot\eta=D^{(1)}(\eta) + \sqrt {2C^{(0)}D^{(2)}(\eta)}\cdot\xi(t),\label{23}
\end{equation}
\noindent
where $\xi(t)$ is a white noise with standard properties
\begin{equation}
\langle \xi(t)\rangle=0,\quad \langle\xi(t)\xi(t^{\prime})\rangle
=\delta(t-t^{\prime}).\label{24}
\end{equation}

For studying the transitions between behaviour regimes of the system let us
use path integral formalism. To achieve this aim we write Langevin
equation in the form of stochastic differential equation
\begin{equation}
d\eta=D^{(1)}(\eta)dt +\sqrt{2C^{(0)}D^{(2)}}dw,\label{25}
\end{equation}
\noindent
where $dw=\xi(t)dt$ represents a
Winner process. This notation allows us to get a new process $y(t)$ with
transition Jacobean
${dy}/{d\eta}=\left(\sqrt{2C^{(0)}D^{(2)}}\right)^{-1}$.
Since we use a white noise then for $y(t)$ the
stochastic differentiation operator can be written
\begin{equation}
dy=\frac{dy}{d\eta}d\eta+ \frac{1}{2} \frac{d^2y}{d\eta
^2}(d\eta)^2.\label{26}
\end{equation}
\noindent
Constructing by such way the evolution equation for $y(t)$ process we
obtain expression for the white noise
\begin{equation}
\xi(t)=\frac{\dot\eta}{\sqrt{2C^{(0)}D^{(2)}}}-\frac{D^{(1)}}{\sqrt
{2C^{(0)}D^{(2)}}}+\frac{1}{2}\left(\sqrt
{2C^{(0)}D^{(2)}}\right)^{\prime}\label{27}
\end{equation}
\noindent
with probability density $P(\xi(t))\propto\exp\left(-\frac{1}{2}\int
\xi^2(t)dt\right)$.    He\-re and further accent means diffe\-ren\-tia\-tion with respect to $\eta$.
Taking into account the relationship between distributions
$P(\eta)=P(\xi)J$, where $J$ is the Jacobean of transition from
$\xi$ to $\eta$ field, according to \cite{14,15}, we
get following expression
\begin{equation}
P(\dot\eta,\eta,t)\propto\exp\left(- \frac{1}{2}
\int {\cal L}dt \right),\label{28}
\end{equation}
\noindent
where Onsager-Machlup function $\cal L$ acts as  Lagrangian in Euclidean
field theory
\begin{equation}
{\cal{L}}=\frac{\dot \eta^2}{2 C^{(0)}D^{(2)}}+ U.\label{29}
\end{equation}
\noindent
Here the effective potential energy $U$ is given by expression
\begin{equation}
U=\left[\frac{D^{(1)}}{\sqrt{2C^{(0)}D^{(2)}}}-\frac{1}{2}\left(\sqrt
{2C^{(0)}D^{(2)}}\right)^{\prime}\right]^2.\label{30}
\end{equation}
\noindent
Thus, the system's kinetics will be characterized by Euler-Lagrange equation
with respect to the function (\ref{29}).

\section{Stationary states, phase diagrams and phase portraits}

At first let us consider a stationary states. Assuming $\dot\eta=0$
in the Euler-Lagrange equation, we obtain
\begin{equation}
\gamma^2\left(f+C^{(1)}\nabla g^2 \right)-
\frac{1}{2}C^{(0)}\gamma\nabla g^2=0,\label{31}
\end{equation}
\begin{equation}
\gamma^3\left(\nabla fg-\nabla gf+2C^{(1)}g^2\nabla^2g\right)
+C^{(0)}g^2\left[(3\nabla g-2)\gamma \nabla \gamma-\gamma^2\nabla
^2g \right]=0.\label{32}
\end{equation}
\noindent
Headway deviation stationary values $\eta_0$, that correspond to the
extremum of the effective potential energy function (\ref{30}),
are determined as solutions of  equation (\ref{31}).
The solution of equation (\ref{32}) gives the point of plateau
appearance in $U$ vs $\eta$ dependence.
Further using definitions for $\gamma(\eta)$, $f(\eta)$, and $g(\eta)$,
we analyze the transitions for stochastic system in details.
The dependence of the order parameter stationary values on noise
intensity and characteristic control parameter is given by equation
\begin{equation}
\eta^4+\eta^2\left(1+2\sigma-\varepsilon-2C^{(1)} \right)-
(\varepsilon-\sigma-2C^{(1)})(1+\sigma)+C^{(0)}=0.\label{33}
\end{equation}
\noindent
In this case the correlation function moments (\ref{6}), which are
determined by equation (\ref{17}), take the following form
\begin{equation}
C^{(0)}=\delta^2,\quad C^{(1)}=\delta^2\tau_{\lambda}.\label{34}
\end{equation}
\noindent
Then from Eqs.(\ref{33}) and (\ref{34}) for the phase diagram
curve, that is the boundary of the existence region of disordered phase
($\eta_0=0$) corresponding to the free traffic, we get
\begin{equation}
\varepsilon=\frac{\delta^2}{1+\sigma}+\sigma-2\delta^2\tau_{\lambda}.
\label{35}
\end{equation}
\noindent
First of all we should note that separation of the low-frequency domain in
noise spectrum results in decrease of the effective time that is necessary
for car to reach a characteristic velocity.

The dependence of the headway deviation stationary values on the
characteristic acce\-le\-ra\-tion/bra\-king time $\eta_0(\varepsilon)$
is the solution of equation (\ref{33}) and is shown in Fig.1a. 
According to Fig.1a  there are zero minimum in the dependence of
effective potential energy on headway deviation, which
corresponds to disordered state (free moving traffic),
and minimum at nonzero stationary value $\eta_0$, which
meets the ordered state (traffic jam formation).
Dotted line in Fig.1a shows that these minimums are separated
by maximum corresponding to unstable state of the system.
Thus, the increase of the noise intensity results in coexistence
of disordered and ordered states inherent in
first-order phase transition.

Phase diagram of the system for different values of the noise
correlation time $\tau_{\lambda}$ is represented in
Fig.2. There are three domains in it. Domain 1 corresponds to disordered
state of the system, i.e. free traffic. Domain 2 is characterized
by coexistence of disordered and ordered states, i.e. free  and
jammed traffic can coexist at such parameters. Last domain 3
corresponds to ordered state of the system. Here only traffic jam exists
(minimum of the effective potential energy) while free traffic is unstable
(maximum of the effective potential energy).

As it is obvious from Fig.2  with increase of the noise correlation time
the domain corresponding to free traffic decreases while domain
of traffic jam increases.

For the consideration of the system kinetics let us use
Euler-Lagrange equation

\begin{equation}
\frac{\partial {\cal L}}{\partial\eta}-\frac{d}{dt}
\frac{\partial {\cal L}}{\partial\dot\eta}
= \frac{\partial R}{\partial\dot\eta},\label{36}
\end{equation}
\noindent
which is supplemented by dissipative function contribution
$R={\dot\eta^2}/{2}$.
Its form is typical for generating
functional method. Taking into account equations (\ref{29}),
(\ref{30}), we find differential equation of second-order

\begin{eqnarray}
\ddot\eta&-&\frac{1}{2}
\frac{\left(D^{(2)}\right)^{\prime}}{D^{(2)}}\dot\eta^2+
C^{(0)}D^{(2)}\dot\eta-
\left[\frac{D^{(1)}}{\sqrt{D^{(2)}}}-C^{(0)}
\left(\sqrt{D^{(2)}}\right)^{\prime}\right]\nonumber\\
&\times&\left[\left(D^{(1)}\right)^{\prime} \sqrt
{D^{(2)}}- D^{(1)}\left(\sqrt{D^{(2)}}\right)^{\prime}-
C^{(0)}D^{(2)}\left(\sqrt{D^{(2)}}\right)^{\prime\prime}
\right]= 0.\label{37}
\end{eqnarray}
Equation (\ref{37}) can be represented as a system of two
first-order differential equations. This reorganization
simplifies the investigation of our model. Such equations set
can be solved using phase-plane method which allows us to
consider the kinetic behaviour of the system on the basis of phase portraits
in the $(\dot\eta;\eta)$ plane.

Phase-plane portraits of such system are shown in Fig.3  for
each domain of phase diagram in Fig.2b for $\tau_{\lambda}=0.2$.
Domain 1 has two singular points: $D$, and unstable node $S$
(Fig.3a). Here singular point $D$ corresponds to free traffic
flow (disordered state). This point has complex character of stability
because phase trajectories at $\dot\eta<0$ converge to $D$
and diverge from it at $\dot\eta>0$ \cite{16}. Coordinate
of singular point $S$ is determined by solution of equation
(\ref{32}). Solutions of equations (\ref{31}), (\ref{32})
for noise intensity $C^{(0)}=10$ are pictured in Fig.4a.
As is obvious from this figure the solution of equation
(\ref{32}) is independent on characteristic
acceleration/braking time $\varepsilon$, i.e. it is shown
by horizontal line. Thus, coordinate of singular point $S$
for all domains of phase diagram may be defined by intersection
of given horizontal line with vertical line for appropriate
$\varepsilon$ value.

Phase portrait for domain 2 of the phase diagram is shown in
Fig.3b. It has four singular points: $D$, unstable nodes $N$
and $O$, and saddle $S$. Similar to the first case singular
point $D$ corresponds to free traffic. Unstable node $N$
corresponds to unstable state of the system, namely, to the
maximum of the effective potential energy. Saddle $S$ is
determined by the solution of equation (\ref{32}).
Unstable node $O$ characterizes the traffic jam formation
(ordered state).

In Fig.3c phase portrait meeting the domain 3 is demonstrated. There
are three singular points: $D$, saddle $S$, and unstable node
$O$. Here point $D$ corresponds to free moving traffic also,
but now phase trajectories which pass through it have closed
form at its vicinity \cite{16}. As previously solution of equation
(\ref{32}) is presented by saddle $S$. Unstable node $O$ corresponds
 to traffic jam.

The above consideration represents supercritical regi\-me of the
traffic jam formation corres\-pon\-ding to second-order phase
transition. However we see that for given system due to
fluctuations of the characteristic accelerati\-on/braking time
free moving traffic and traffic jam can coexist that inherent
in the first-order phase transition (sub\-critical re\-gi\-me). In
addition the phase portraits (Fig.3a-c) show that singular point
$O$ meeting the  traffic jam state is unstable. Thus, we should
consider subcritical regime of the traffic jam formation
which is a true reason of self-organization and analogous
to the first-order phase transition.

For this let us assume that the relaxation time of
headway deviation is the function  $t_{\eta}(\eta)$
increasing with $\eta$ from initial value
$t_{\eta}(1+k)^{-1}$, $k>0$ to the final one $t_{\eta}$, and is
determined by the simplest approximation

\begin{equation}
\frac{t_{\eta}}{t_{\eta}(\eta)} = 1 +
\frac{k}{1 +\left(\eta/\eta_t \right)^{2}},\label{38}
\end{equation}
\noindent
where $0<\eta_t<1$, $k$ and $\eta_t$ are dispersion constant and scale.
Then doing the same operations as in the supercritical case we obtain
equation (\ref{11}), but now $\gamma(\eta)$ and $f(\eta)$ are defined by
the following expressions:

\begin{eqnarray}
\gamma(\eta) &=& 1+\sigma + k \frac{1 -\eta^2/\alpha^2}
{\left( 1+\eta^2/\alpha^2 \right)^2} + \eta^2,\nonumber\\
f(\eta)&=&\left[\varepsilon-\sigma\left(1 +
\frac{k}{1 + {\eta^2}/{\alpha^2}}\right) \right]\eta-
\left(1+\frac{k}{1+{\eta^2}/{\alpha^2}}\right)\eta^3,\label{39}
\end{eqnarray}
\noindent
where $\alpha\equiv {\eta_t}/{\eta_m}$.
Further  Fokker-Planck equation, Lan\-ge\-vin equation, and
Onsager-Machlup function are determined by the similar manner.
However, insertion of Eqs.(\ref{39}) into Eq.(\ref{31})
gives the following expression for the stationary values of
headway deviation:
\begin{equation}
\eta^4 d+\eta^2\left[d(m+\sigma)-\varepsilon-2C^{(1)}\right]
-m\left(\varepsilon -\sigma d+2C^{(1)} \right)+C^{(0)} = 0,\label{40}
\end{equation}
\noindent
where
\begin{eqnarray}
d&=&1+\frac{k}{1+\eta^2/\alpha^2},\nonumber\\
m&=&1+\sigma+k\frac{1 -\eta ^{2}/\alpha ^{2}}
{\left(1+\eta ^{2}/\alpha ^{2}\right)^2}.\label{41}
\end{eqnarray}

Equations (\ref{34}), (\ref{40}) and (\ref{41}) define the boundary of the
existence domain of disordered phase ($\eta_0=0$) in phase diagram
\begin{equation}
\varepsilon=\frac{\delta^2}{1+\sigma+ k}+\sigma(1+k)-
2\delta^2\tau _{\lambda}.\label{42}
\end{equation}
\noindent
According to this the separation of the low-frequency 
region in the noise spectrum leads to decrease of the 
time needed for car to reach a characteristic velocity.

Dependence $\eta_0(\varepsilon)$, that is a solution of equation
(\ref{40}), is pictured in Fig.1b. This figure show 
that increase of the noise intensity $C^{(0)}$
results in appearance of two stationary states corresponding to
minimums of the dependence of effective potential energy on 
headway deviation $\eta $. Hence, we can conclude that two
stationary values of headway deviation from optimal value exist. At
these values traffic jam and congested traffic can be formed
\cite{10}. The smaller value corresponds to the metastable state
(dashed curve) while lager value (solid line) meets the stable ordered
state of the system. These states are divided by unstable state
(dotted line) corresponding to maximum of the effective potential energy.

Phase diagrams of the system is shown in Fig.5 for different
values of the noise correlation time $\tau_{\lambda}$. Here domain
1 corresponds to disordered state of the system, in other words,
only one minimum  of the dependence of effective potential
energy on headway deviation exists meeting zero value. This minimum
characterizes the free traffic. In domain 2 disordered and
ordered states coexist, so that in addition to zero minimum
the minimum appears meeting the nonzero value of headway
deviation. This minimum corresponds to traffic jam formation.
Domain 3 meets the most complex form of effective potential
energy. Here the metastable and ordered states of the system coexist.
Me\-ta\-stab\-le state appearance is caused by displacement of  zero
minimum (disordered state) along the axis of headway
deviation of the dependence of effective potential energy.
It implies that in this domain traffic jam can appear
at two different values of headway deviation
from optimal value. And at the same time zero value of headway
deviation corresponds to maximum of effective potential energy,
i.e. free traffic becomes unstable. Domain 4 characterizes the
ordered state of the system, i.e. traffic jam at a given headway
deviation value and unstable free traffic. Last domain 5
describes metastable system's state and is located near the
boundary between domain 1 and 3. At crossing the boundary
from domain 1 to domain 5 transition of the displacement
type from disordered to metastable state occurs. It corresponds
to appearance of only one minimum of the effective
potential energy and means the opportunity of congested traffic
formation at small value of headway deviation. Free traffic
is unstable here too.

Let us now analyze kinetics of the system using Euler-Lagrange
equation (\ref{36}). Phase-plane portraits 
of the system for four domains, that are shown in phase
diagram for $\tau_{\lambda}=0.2$ (Fig.5b), are demonstrated in Fig.6.
Domain 1 is characterized by the presence of three singular points:
$D$, stable focus $F$ and saddle $S$ (Fig.6a). Here singular
point $D$ corresponds to free traffic and has the same
complex character with singular point $D$ for
supercritical regime. Coordinates of the stable focus $F$
and saddle $S$ are determined by the solutions of equation
(\ref{32}) with taking into account Eqs.(\ref{39}). Solutions of
Eqs.(\ref{31}), (\ref{32}) for subcritical regime for
noise intensity $C^{(0)}=10$ are shown in Fig.4b. As is
obvious from the figure the solutions of equation (\ref{32}) are
independent on characteristic acceleration/braking time $\varepsilon$.
Consequently, we may obtain the coordinates of singular points $F$
and $S$ by intersection of corresponding horizontal lines with
vertical lines for appropriate $\varepsilon$ value. Thus the
coordinates of singular points $S,S^{\prime},F,F^{\prime}$
can be defined for all phase diagram domains.

Phase portrait of the system for domain 2 of phase diagram is
shown in Fig.6b. There are four singular points: $D$, saddles
$N$ and $O$, and stable focus $F^{\prime}$. As in the Fig.6a
singular point $D$ corresponds  to free traffic, but now phase
trajectories which pass through it have closed form at its
vicinity \cite{16}. Point $F^{\prime}$ is determined by the solution
of equation (\ref{32}). Saddle $O$ characterizes the traffic
jam formation. Saddle $N$ corresponds to unstable state of the
system, namely, to maximum of the dependence of effective
potential energy on headway deviation $\eta$.

In Fig.6c phase portrait for the most complex domain 3 of phase diagram is
demonstrated. The maximal number of singular points is placed here: $D$,
saddles $M$, $N$ and $O$, stable focus $F$, and stable node $F^{\prime}$.
As in the Fig.6a coordinates of points $F$ and $F^{\prime}$ are
determined by solutions of equation (\ref{32}). Singular points
$N$ and $O$ have the same meaning as in Fig.6b. Saddle $M$
corresponds to metastable state of the system, namely, to the
formation of congested traffic.

Phase portrait in Fig.6d corresponds to domain 4 of phase diagram. As
is shown above the solutions of equation (\ref{32}) are represented
by saddle $S^{\prime}$ and stable node $F^{\prime}$. Saddle $O$
corresponds to jammed traffic.

\section{Discussion}

The above consideration shows that for supercritical regime, which
corresponds to the second-order phase transition, due to
fluctuations of the characteristic acceleration/bra\-king time
free and jammed traffic can coexist. As is known such a
picture is inherent in first-order phase transition. For subcritical
regime, which meets the first-order phase transition, due to
regulation of fluctuations intensity of characteristic
acceleration/braking time a given system can be passed from free
traffic regime to traffic jam formation. The last state can
appear at  different values of headway deviation between transport
facilities. State of the system corresponding to the smaller value
of headway deviation may be represented as a congested traffic regime.
However, the correct answer can be defined using additional investigation of
dynamical car characteristics, but such study is out of our consideration
framework.

It is shown that transition to low-fre\-quen\-cy
fluctuations of acceleration/braking time (increase of $\tau_{\lambda}$)
leads to oppose to free moving traffic regime bounding its
domain.  It is worthwhile to note that for subcritical regime
of self-organization the singular points, that correspond to
effective potential energy extremum, are saddles. Such situation
is typical for description of stochastic system by Euclidean
field theory method \cite{17}.

\bigskip
\newpage
\begin{center}
{\bf Captions}\\
to the paper
"Jamming transition with fluctuations of characteristic
acceleration/braking time within Lorentz model"
by A.V. Khomenko, D.O. Kharchenko, O.V. Yushchenko
\end{center}

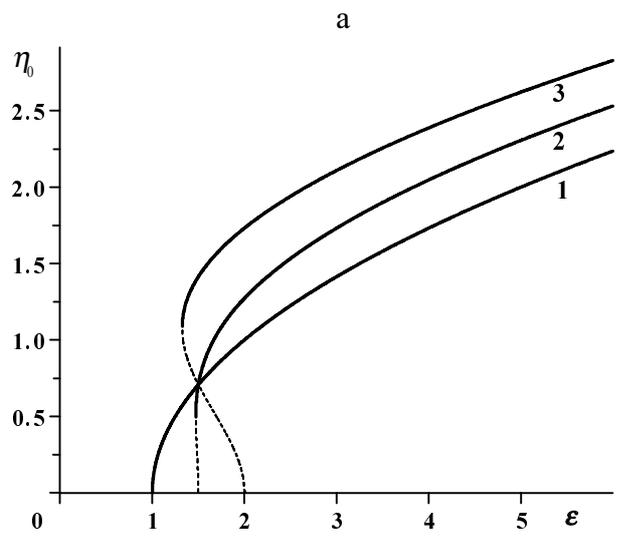
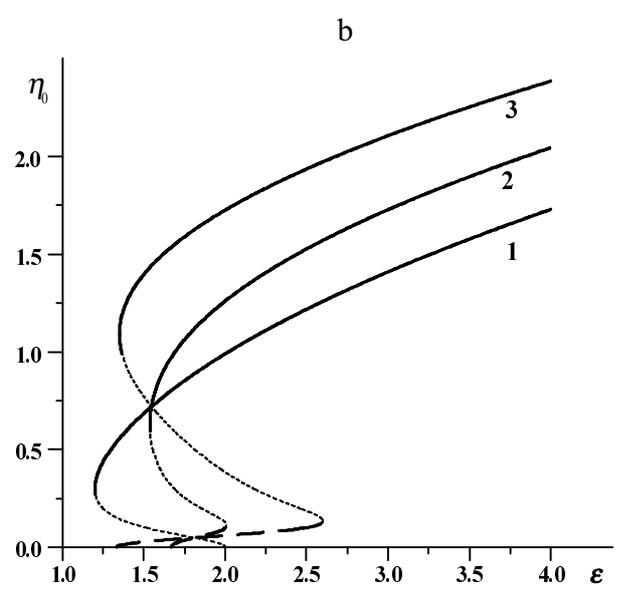
Fig.1 -- Dependence of the stationary value of headway deviation $\eta_0$
on cha\-racte\-ris\-tic ac\-ce\-le\-ra\-tion/braking time
$\varepsilon=\tau_0/\tau_c$ at different noise intensities $C^{(0)}$
of $\varepsilon$ (curve 1 corresponds to $C^{(0)}=0$, 2 -- $C^{(0)}=5$,
3 -- $C^{(0)}=10$) for $\sigma=1$ and  $\tau_{\lambda}=0.2$:
(a) for supercritical regime; (b) for subcritical regime at
$\alpha = 0.1$ and $k=1$.
\bigskip

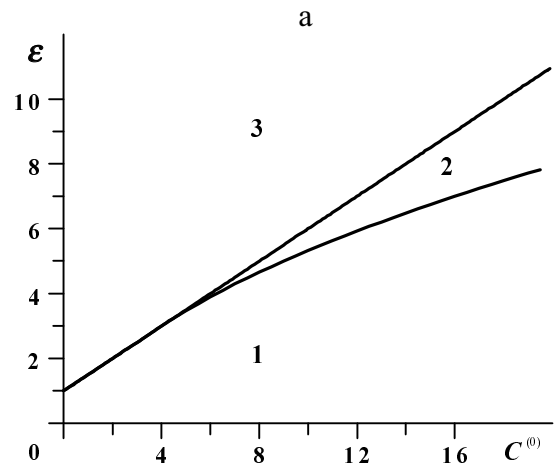
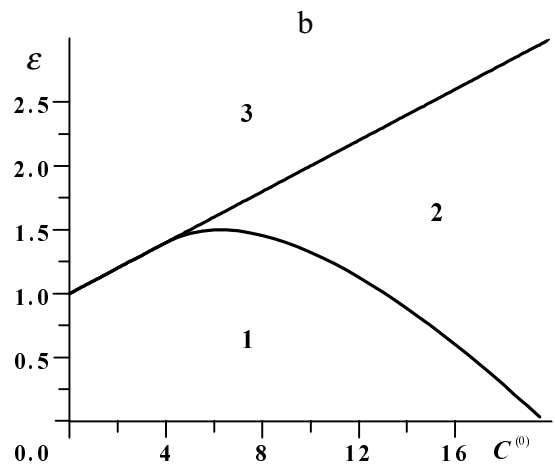
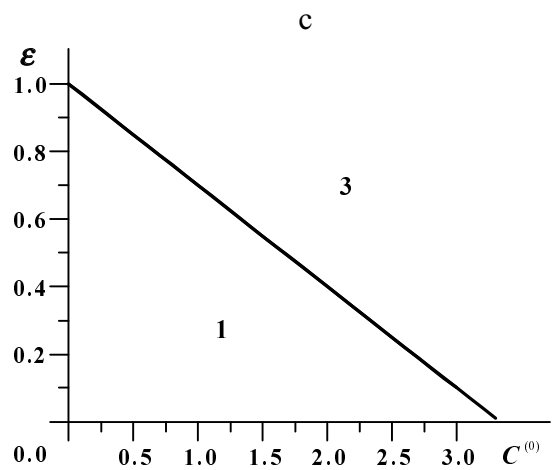
Fig.2 -- Phase diagrams of the system at supercritical regime
for different values of noise correlation time 	at
$\sigma=1$: (a) $\tau_{\lambda}$=0.0; (b)
$\tau_{\lambda}=0.2$; (c) $\tau_{\lambda}=0.4$. Domain 1 corresponds to
free moving traffic, 2 -- traffic jam and free traffic, 3 -- traffic jam.
\bigskip

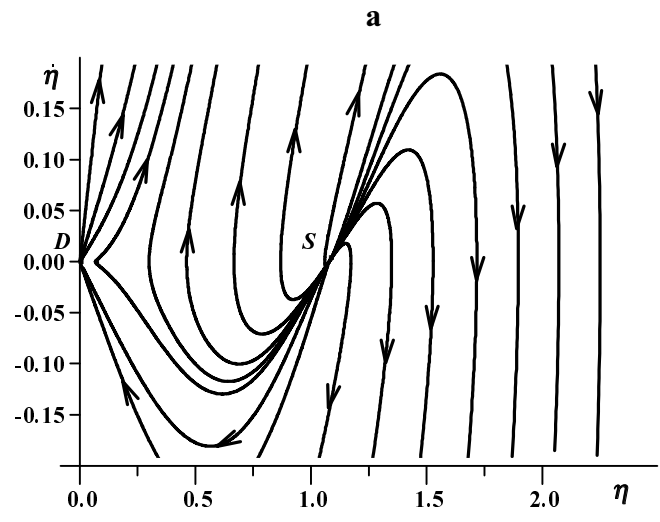
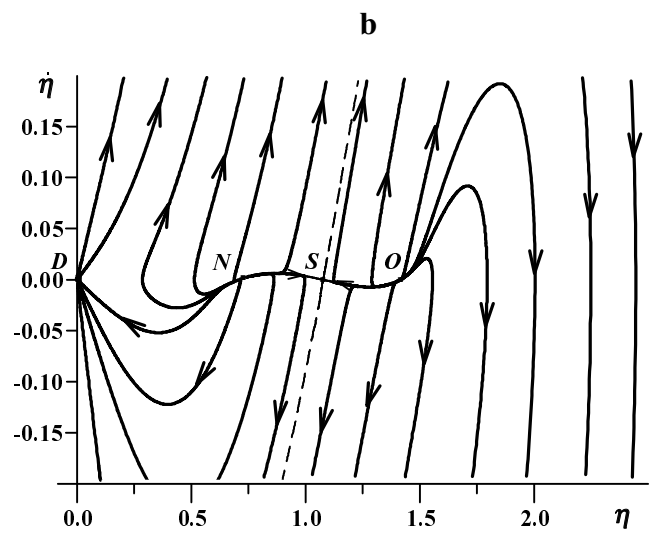
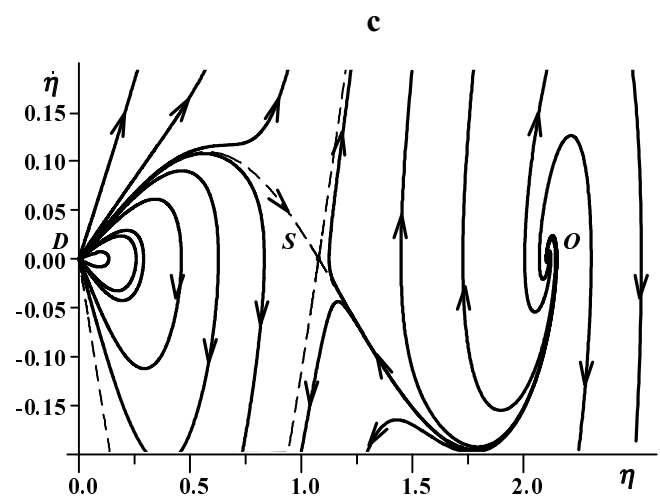
Fig.3 -- Phase portraits of the system at supercritical regime for
different domains of phase diagram at  $C^{(0)}=10$,
$\sigma=1$, and $\tau_{\lambda}=0.2$: (a) domain 1 at
$\varepsilon=0.5$; (b) 2 -- $\varepsilon=1.5$; (c) 3 -- $\varepsilon=3$.
\bigskip

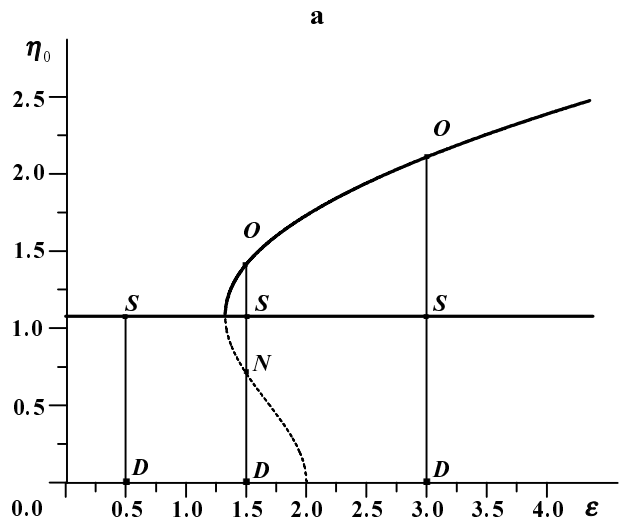
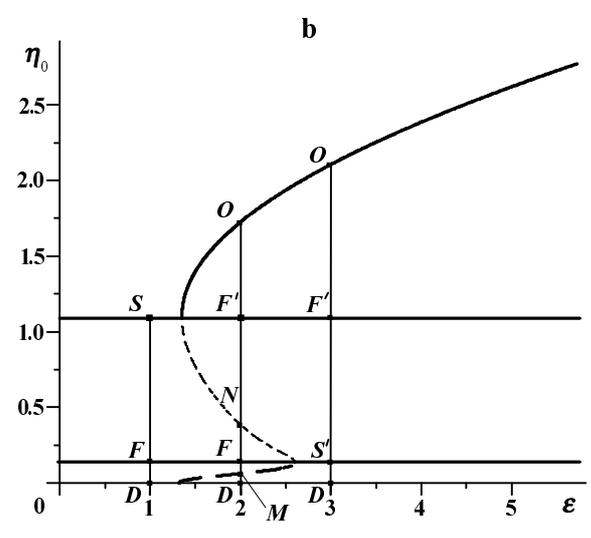
Fig.4 -- Dependence of solutions $\eta_{0}$ of equations
(\ref{31}) and (\ref{32}) for stationary states on
characteristic acceleration/braking time
$\varepsilon={\tau_0}/{\tau_c}$ at $C^{(0)}=10$,
$\tau_{\lambda}=0.2$, and  $\sigma=1$: (a) for supercritical regime;
(b) for subcritical regime at $\alpha=0.1$, $k=1$.
\bigskip

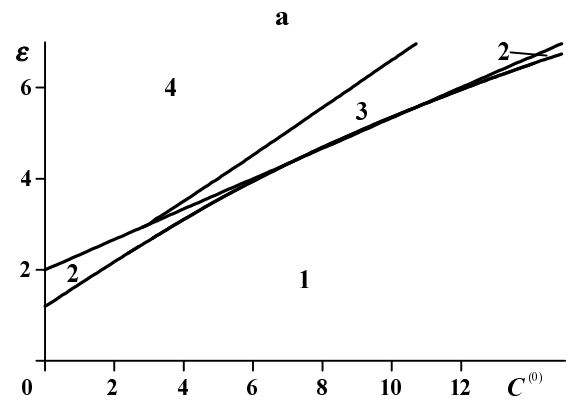
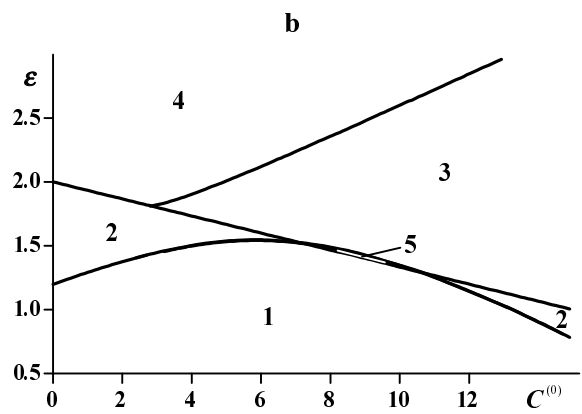
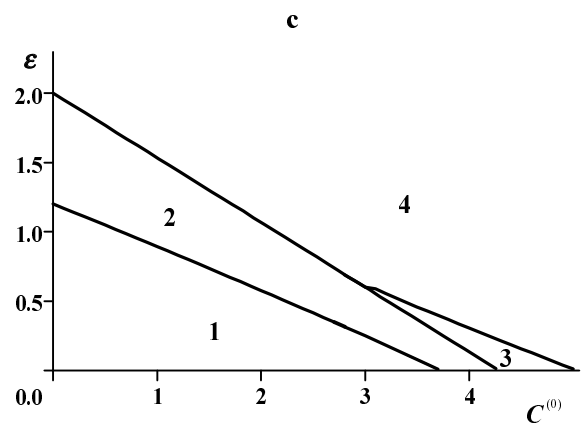
Fig.5 -- Phase diagrams of the system at subcritical regime
for different values of noise correlation time 	at
$\alpha=0.1$, $\sigma=1$, and $k=1$: (a) $\tau_{\lambda}$=0.0; (b)
$\tau_{\lambda}=0.2$; (c) $\tau_{\lambda}=0.4$. Domain 1 corresponds to
free moving traffic, 2 -- traffic jam and free traffic, 3 --
metastable congested traffic and traffic jam, 4 -- traffic jam,  and 5 --
metastable congested traffic.
\bigskip

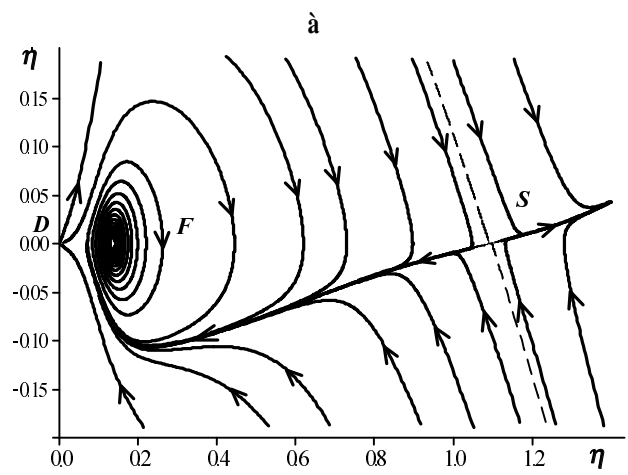
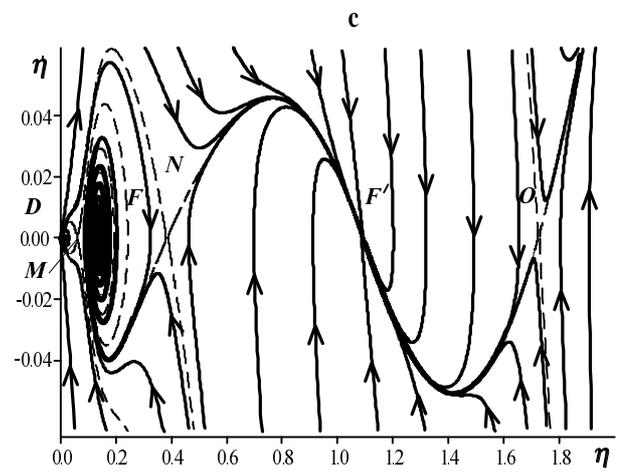
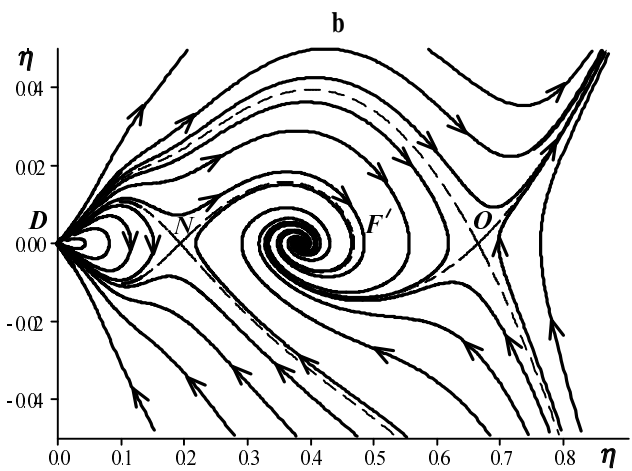
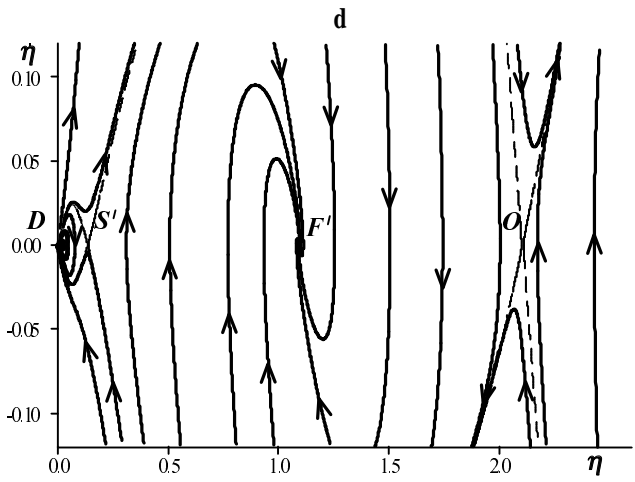
Fig.6 --  Phase portraits of the system at subcritical regime for
different domains of phase diagram at $\alpha = 0.1$, $\sigma=1$,
$k=1$, and $\tau_{\lambda}=0.2$: (a) domain 1 at $C^{(0)}=10$,
$\varepsilon=1$; (b) 2 -- $C^{(0)}=2$, $\varepsilon=1.5$;
(c) 3 -- $C^{(0)}=10$, $\varepsilon=2$;
(d) 4 -- $C^{(0)}=10$,  $\varepsilon=3$.


\begin{thebibliography}{00}
\bibitem{1} T. Nagatani, Phys. Rev. E {\bf 58},  4271 (1998).
\bibitem{2} R. Mahnke, J. Kaupu\v{z}s,  Phys. Rev. E {\bf 59},  117 (1999).
\bibitem{3} D. Chowdhury, L. Santen,  A. Schadschneider,
 Physics Reports {\bf 329}, 199 (2000).
\bibitem{4} T. Nagatani,  Phys. Rev. E {\bf 61}, 3534 (1999).
\bibitem{5} E. Ben-Naim, P.L. Krapivsky,  Phys. Rev. E {\bf 59}, 88 (1999).
\bibitem{6} L. Neubert, H.Y. Lee, M. Schreckenberg, J. Phys. A  {\bf 32},
6517 (1999).
\bibitem{7}  W. Horsthemke and R. Lefever,  {\it Noise-Induced Transitions.
Theory and Applications in Physics, Chemistry, and Biology}
(Springer, Berlin, 1984).
\bibitem{8} H. Haken, {\it Synergetics. An Introduction.  Nonequilibrium
Phase Transitions and Self-Organization in Physics, Chemistry, and
Biology}, 3rd edn. (Springer, Berlin, 1983).
\bibitem{9} A.I. Olemskoi, A.V. Khomenko,  Sov. Phys. JETP {\bf 83},
1180 (1996).
\bibitem{10}  A.I. Olemskoi, A.V. Khomenko,  Phys. Rev. E {\bf 63},
 036116 (2001).
\bibitem{11} N.G. van Kampen, {\it Stochastic Processes in Physics,
and Chemistry} (North-Holland, Amsterdam, 1981).
\bibitem{12} V.E. Shapiro,  Phys. Rev. E {\bf 48}, 109 (1993).
\bibitem{13} D.O. Kharchenko,  Ukr. Phys. Jorn. {\bf 44}, 647 (1999).
\bibitem{14} G. Repke, {\it Nonequilibrium statistical mechanics}
(Mir, Moscow, 1990).
\bibitem{15} P. Arnold, Phys. Rev. E {\bf 61}, 6099 (2000).
\bibitem{16}  N.N. Bautin, E.A. Leontovich,
{\it Metody i priemy  kachestvennogo issledovaniya dinamicheskikh
sistem na ploskosty} ({\it Methods and techniques of qualitative
investigation of dynamical systems in the plane})
(Nauka, Moscow, 1976) (in Russian).
\bibitem{17}  A.S. Mikhailov, A.Yu. Loskutov, {\it Foundations
of Synergetics II} (Springer-Verlag, Berlin-Heidelberg, 1996).
\end{thebibliography}
\end{document}